# Feature Subset Selection for Software Cost Modelling and Estimation

Efi Papatheocharous[1], Harris Papadopoulos[2] and Andreas S. Andreou[3],

[1] University of Cyprus, Department of Computer Science,
75 Kallipoleos Street, P.O. Box 20537, CY1678 Nicosia, Cyprus
efi.papatheocharous@cs.ucy.ac.cy

[2] Frederick University, Computer Science and Engineering Department,
7 Y. Frederickou Street, Palouriotisa, 1036 Nicosia, Cyprus
h.papadopoulos@frederick.ac.cy

[3] Cyprus University of Technology, Department of Electrical Engineering and Information Technology, 31 Archbishop Kyprianos Street, 3036 Lemesos, Cyprus
andreas.andreou@cut.ac.cy

**Abstract**

Feature selection has been recently used in the area of software engineering for improving the accuracy and robustness of software cost models. The idea behind selecting the most informative subset of features from a pool of available cost drivers stems from the hypothesis that reducing the dimensionality of datasets will significantly minimise the complexity and time required to reach to an estimation using a particular modelling technique. This work investigates the appropriateness of attributes, obtained from empirical project databases and aims to reduce the cost drivers used while preserving performance. Finding suitable subset selections that may cater improved predictions may be considered as a pre-processing step of a particular technique employed for cost estimation (filter or wrapper) or an internal (embedded) step to minimise the fitting error. This paper compares nine relatively popular feature selection methods and uses the empirical values of selected attributes recorded in the ISBSG and Desharnais datasets to estimate software development effort.

1. Introduction

Software cost estimation is one of the basic project management processes carried out to support efficiently resource allocation activities. Basically it involves approximating the person-months required for developing software systems. The development activities taken into consideration in cost estimation usually start after product specification and continue until product implementation and delivery. Accurate estimates enable project managers to effectively react to contingencies and reallocate, if necessary, budget and effort. Managers and researchers in the industry and academia respectively have been striving for decades to improve the accuracy and consistency of effort estimations. To achieve this, several techniques have been proposed with the ultimate goal to have successful project completion and develop high quality software with the lowest possible costs [25].
Finding effective techniques to estimate the overall development costs, especially from the planning phases of a project, can offer a competitive advantage for software companies and managers [18]. This competitive advantage leads to risk minimisation, reduction of resource misallocation and enhancement of the manager's ability to deal with budgeting, planning and controlling the project processes and assets. The effectiveness of a software cost estimation technique is typically measured in terms of accuracy [42]. However, effectiveness in a particular technique may be measured in terms of producing the intended or expected result performing in the best possible manner, without wasting time on useless calculations using irrelevant or redundant information. Therefore, some important questions to ask are: Which information, and in particular, which project features should we use as inputs for a certain cost modelling technique? Can we reduce the complexity of calculations for a given technique? Should we use all of the project attributes available at the time of the estimate, or a subset that appears to be most relevant? This paper will attempt to address the above questions.

Software cost estimation models make use of a wide range of project attributes, often called cost drivers, for approximating development effort. Cost drivers relate to the actual size (code length or function points), complexity, duration of the project, experience of the people involved in developing the system, etc.; all these are attributes that are not easy to quantify with accuracy. This is the primal reason for which many cost models proposed in the relative literature often utilise expert judgement by taking into consideration the opinion of expert cost estimators [19]. Also, usually, practitioners calibrate regressions [26, 22] based on empirical databases which involves building and evaluating models using a set of historical projects to achieve better cost approximations. In essence, practitioners make use of known project data and the observed relationship between the independent variables and the dependent variable (effort) to make estimates of the latter for new projects, given only the new values of their independent variables. Moreover, machine learning techniques [31, 47, 20] have recently gained ground in literature, according to recent surveys [25]. Despite the vast research background, estimating accurately the effort for software projects is still considered a difficult task [10, 36, 25]. The main reason that cost estimations are usually inaccurate is that some cost drivers affecting development effort relate to highly subjective characteristics, are not yet adequately investigated, defined or understood by researchers, and the various cost drivers exhibit complex inter-relationships that cannot be easily modelled. In addition, due to the fact that software development is driven by a large number of factors that are consequently used in cost estimation models, for example the popular COCOMO uses 16 cost drivers [6] and COCOMO II uses 23 cost drivers [8], it can result in strong collinearity, heteroscedasticity and unstable prediction accuracy [33, 37]. Finally, recent studies report that the quality and appropriateness of the datasets utilised in most effort estimating techniques are key factors to obtaining better results [30].

Previous investigations emphasised on the ability of Artificial Neural Networks (ANN) to signify interactions between cost drivers and the actual effort required to develop a set of historical projects [40]. Garson's algorithm [17] was used on the best trained networks to evaluate the degree of influence of project features when successful effort predictions were obtained so as to select subsets of features with the highest contribution to the result. This paper extends previous work on the topic [40, 41] and leverages various feature selection methods to explore the possibility of reducing the cost drivers required in software cost estimation and either retain or improve accuracy performance. The main goal of this work is to assess the appropriateness of software features in cost modelling and estimation, which will subsequently lead to a better understanding of the features that may be used for achieving high estimation performance, or to considering their relevance, redundancy or irrelevancy. Besides feature interpretation, we are also concerned with the number and type of features selected by various approaches. In order to evaluate the quality of the selected feature subsets, and therefore their appropriateness, we compare their accuracy performance to that of the full feature set using a specific estimation technique, namely dual variables Ridge Regression (RR), on two real-life datasets, ISBSG and Desharnais. By selecting the appropriate subsets of features we additionally aim in reducing the complexity of software cost estimation models and subsequently eliminating the need to measure and manage large volumes of project cost drivers/parameters.

Notably, using search approaches in high dimensional datasets, i.e., datasets with a large number of available attributes, for removing the most irrelevant and redundant attributes is a tedious task. The exhaustive search of all possible subsets of features is impractical and extremely time-consuming. Based on our previous experience, in data-intensive estimation methods, such as ANN and RR, a significant task is the pre-processing performed to remove irrelevant data which may lead to less complex and equal or more accurate effort approximations. Moreover, accurate software cost estimations are not the result of a purely blind process that takes any number of inputs found in empirical software databases and outputs work effort, but seem to be influenced by various factors to different degrees. Therefore, our investigation was performed under the assumption that some cost drivers are more informative than others and we attempted to select the satisfactory or optimal number of features.

The rest of this paper is organised as follows: Section 2 gives an overview of software cost estimation models and related research work. Section 3 presents the feature selection approaches utilised in the experiments. Section 4 describes the evaluation methods and accuracy measures used in the experimental process. Section 5 discusses the main results obtained, while section 6 summarises the conclusions and suggests future research steps.

2. Literature Overview

The software cost estimation models proposed in the literature may be divided into three main categories namely, (i) expert judgment, (ii) algorithmic and (iii) machine learning. Expert judgment models rely purely on the experience and knowledge of one or more experts [19]. Algorithmic models [7, 44] involve the utilisation of analytical or statistical equations relating software project cost to a number of input parameters. The input parameters, also called predictor metrics, that is, metrics that can be measured during one phase of a project and can be utilised to define the value of another metric in a later phase, are commonly used to create parametric models. Well-known traditional parametric models are: Boehm's COCOMO [6, 8], Albrecht's FPA [2, 1] and Putnam's SLIM [43, 44, 45]. These models rely on statistical regression-based techniques utilising various project cost factors. On the contrary, machine learning makes no or minimum assumptions regarding the form of attributes used as prediction metrics and combines notions of soft computing to form estimators. Such approaches deal with the challenge of producing reliable and accurate estimations even across development environments since they are able to adjust to the conditions under study.

In recent years, the diversity and growth of cost estimation approaches have increased and a tendency towards alternative approaches like, genetic programming, linear programming, soft computing, fuzzy logic modelling etc. has been reported [25]. Such approaches of advanced intelligence, even though learn by example, are able to generalise the knowledge obtained, adapt and exceed statistical approaches that are more sensitive and influential to the accuracy of software metrics [18]. These sophisticated and intelligent models may approximate effort more adequately, and if used in conjunction with an appropriate search technique, they may assist in selecting the most suitable feature subset for cost estimation among a large number of factors. The selection of the most suitable feature subset in each model may: (i) minimise the model's training time, (ii) reduce input data dimensionality and (iii) improve the understanding of the explanatory power of the prediction metrics usually used in software cost estimation.

Many researchers questioned the necessity of a large number of features involved usually in cost estimation techniques and moreover, investigations showed that in most cases redundant features could be eliminated. Feature selection involves finding the optimum number and subset of features that provide the most accurate prediction. Some earlier research work involved discriminant analysis and correlations for dimensionality reduction [49], exhaustive search of features [47], hill climbing and forward sequential selection [28].

Menzies et al. [35] proposed that feature subset selection should be regularly carried out in software cost estimation. The authors utilised a Wrapper linear regression and ranked the frequency with which each feature was selected by the algorithm to form groups of features with the same rankings. Features were then removed in ranked order. Using the best features the authors found that effort prediction results were always improved, even though for datasets with too many projects, the improvement was very small. In [12] stratification and feature subset selection was performed using a Wrapper and linear regression. Experimental results showed that the combination of row and column pruning can improve dramatically effort estimation results especially in small datasets. Investigating the same data with the previously mentioned work, Jalali et al. [24] found relatively small improvements from row pruning and presented results that disagree with the previous research findings in [12]. In fact, the authors suggested that feature selection, i.e. column pruning alone (without row pruning) is considerably useful at reducing the mean and standard deviation of the error obtained by the models examined. In [13] the accuracy of the COCOMO was improved with a Wrapper technique that was used to identify the most promising features for the model. Even though Wrappers were popular in previous research, they yield accurate results at the cost of high computational power and low generalisation of the feature subset selected to other models.

In [4] an extensive feature weight search in analogy-based software cost estimation was proposed using the weighted Euclidean distance, with respect to obtaining an optimal (lowest) error value, which led to increasing the estimation accuracy and reliability of the method. In addition, Genetic Algorithms (GA) were used to assign proper weights to features, and three different heuristics were proposed to increase the estimation performance [21]. The authors identified that encouraging results were obtained when GA were applied coupled with an analogy-based method, improving the accuracy of effort estimates.

Other Feature Subset Selection (FSS) algorithms were investigated for increasing accuracy performance in analogy-based software effort estimation models [5]. The approach, based on fuzzy logic, was able to perform as well as other algorithms, such as hill climbing, forward subset selection and backward feature elimination. Keung et al. [27] proposed Analogy-X to select the best set of features and validate the appropriateness of estimation by analogy using Mantel's Correlation. The authors applied a method similar to stepwise regression analysis to perform sensitivity analysis, detect significant relationships, extract features and locate abnormal data. However, the method's main limitation was its inability to handle categorical values. Li et al. [32] discussed how most FSS methods in analogy-based estimations are implemented as Wrappers and taking into account the advantages of Filter approaches such as selecting more appropriate features than wrappers that merely optimise the error measure, they proposed a hybrid Wrapper and Filter algorithm. Their results showed that the proposed method was an even more effective feature selector that could overcome some of the limitations and computational costs of other techniques proposed in the field.

Taking previous related work into consideration, in this paper we start from the assumption that some predictor metrics are more informative than others and we perform a comparative study of different FSS methods to reduce the number of attributes and identify the most relevant attributes in relation to the model's output. In previous work Ridge Regression (RR) has been proven promising to drive accuracy performance to higher levels [39, 33]. While the focus in [39] was on deriving prediction intervals for effort estimation and in [33] on the multi-collinearities of datasets which were found to lead to unstable regression coefficients, this work investigates several FSS and develops an algorithm based on RR that selects the optimum cost attributes of two widely known real datasets so that the model applied for software cost estimation is improved and simplified. The datasets selected for experimentation are commonly used in the cost estimation literature, while one of the two has a multi-dimensional form, contains a plethora of software projects and a large number of categorical and multi-valued attributes. Finally, it is regarded highly difficult to extract the more informative features from datasets of this complexity and size (as the one used in our experiments), and also time-consuming to locate and gather the most important (in terms of influence) cost drivers for estimating development effort.

## 3. Feature Selection Approaches

The central aim of feature selection approaches is the identification of the most appropriate subset of features for a specific problem. Typically, feature selection approaches may belong to one of three categories: Wrappers, Filters or Embedded algorithms. Wrappers utilise the machine learning algorithm as a black box to rank feature subsets under examination according to their accuracy prediction. Filters, as a pre-processing step, will filter the feature subsets independently of the machine learning algorithm. Embedded methods are included as part of a specific machine learning technique and through training they provide subset selection for the specific technique.

In this work various popular FSS approaches have been selected and applied to the problem of software cost estimation. As expected, each technique resulted in the selection of a different cost driver subset, with some cost drivers appearing in the majority of subsets and others appearing very rarely or not at all. This triggered us to examine both the quality of the different subsets as well as their similarities.

For each feature selection approach performed we randomly partitioned the datasets used in the experiments into training and testing sets 10 times, each time allocating 80% of the total projects to the training set and the remaining 20% to the testing set. The training samples were used to select the optimum features with each technique and then evaluation of the optimum features was carried out with the testing samples. This process was followed to ensure randomness of the samples used to train the algorithms and that an independent sample set (i.e., that was not known during the training) was used for the evaluation of the selected features.

In the rest of this section the feature selection approaches utilised in the experiments of Section 5 are described.

### 3.1 Stepwise Regression

Regression analysis is the most widely applicable method in literature and is used to capture the relationship of the features and effort in the form of a linear function. Each set {$(x_1,z_1)$, ..., $(x_n,z_n)$} represents a sample of projects, where $x_i \in \mathcal{R}^n$ is a vector consisting of the independent (input) variables and $z_i \in \mathcal{R}$ is the dependent variable for each project $i$. We assume a model of the form:

$$z_i = f(x_i) + \varepsilon_i \qquad (1)$$

where the errors $\varepsilon_i$ are independent and have a zero mean. Our goal is to find the coefficients $\beta_0$ and vector $\beta_1$ of the regression linear function:

$$f(x_i) = \beta_0 + \beta_1 x_i \qquad (2)$$

that minimise the sum of squares error. In case the relationship between the dependent and independent variables is not linear we assume that a simple transformation such as logarithmic can be used to estimate a model of the form:

$$z_i = \beta_0 + \beta_1 x_i + \varepsilon_i. \qquad (3)$$

Matlab's function *stepwisefit* was utilised as a Filter approach to select subsets of features. Stepwise Regression was used both in a forward and a backward manner aiming to find the right number of features that provide the best fit for a given response variable. In the forward manner we begin with an empty set of features and in the backward manner with a full feature set and we add or remove features respectively, if the prediction is improved. The function was used in a systematic manner for adding variables (in the case of forward selection), called Forward StepWiseFit (FSWF) and for removing variables (in the case of backward elimination), called Backward StepWiseFit (BSWF).

The method is applied by establishing an initial multilinear model using either all the available attributes or an empty set of variables. Then, the method adds or removes variables to the model according to the explanatory power, added or removed in each case, by examining common statistics for significance in regressions. This means that each time the model changes, we calculate the *p-value* of the *F-statistic* to test the model with or without the 'candidate' variable. The candidate variable is the variable that in each step is assessed in order to be added or removed.

Particularly, if a variable is not currently included in the model, the null hypothesis is that this variable would have a zero coefficient if it was added to the model. However, if there is sufficient evidence to reject the null hypothesis, then the variable is added to the model. Conversely, if a variable is currently in the model, the null hypothesis is that the variable has a zero coefficient. If there is insufficient evidence to reject the null hypothesis, then the variable is removed from the model.

The method continues with the following steps:
- *Step 1*: Fit the initial model.
- *Step 2*: Add the variable that has the smallest *p-value*, from the candidate variables that do not currently participate in the model and have *p-values* less than an inclusion threshold (that is, it is unlikely that they would have zero coefficient if they were added to the model) and repeat this step; otherwise, go to step 3.

- *Step 3*: Remove the variable that has the largest *p-value*, from the variables included in the model that have *p-values* greater than a removal threshold (that is, it is unlikely that the hypothesis of a zero coefficient can be rejected) and proceed to step 2; otherwise, end.

The method terminates when no single step improves the model [16]. The maximum *p-value* for a predictor to be added was set to 0.05 and the minimum *p-value* to be removed was set to 0.10.

3.2 Garson's Algorithm on Artificial Neural Networks (ANN)

One of the primary applications of Artificial Neural Networks (ANNs) involves forecasting a dependent variable $z_1$ from a given set of independent variables $\{x_1,..., x_n\}$. ANNs are non-linear, model-free and alternative to traditional statistical methods. ANNs consist of basic computational elements called neurons organised in groups forming layers. Certain types of neurons organised in multiple layers form the Multi-Layer Perceptron (MLP) [34] which is one of the most popular networks. A simple MLP ANN is shown in Figure 1.

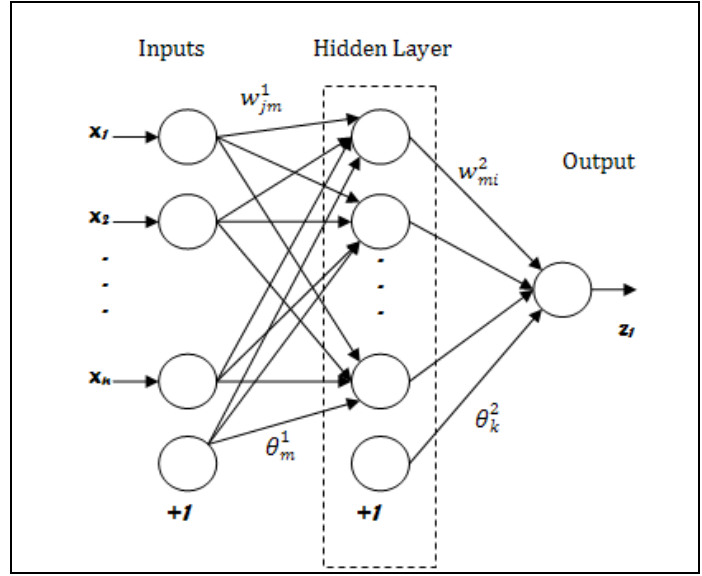

Figure 1: A Feed-forward Multi-Layer Perceptron (MLP) ANN

The number of neurons in the input (first) layer of the ANN is equal to the number of attributes used as independent variables. The last layer is the network output. During learning the weights are adjusted so that the error between the actual and predicted values for the dependent variable is diminished. Each neuron uses the respective input vectors, the weights and bias as specified in equation (4) which shows an example of how the outcome of hidden node $x_1^1$ is estimated. The inputs ($\{x_1,..., x_n\}$) fed in the respective neuron are multiplied by their respective weights ($\{w_{11}^1,..., w_{n1}^1\}$) and adding the bias value ($\theta_1^1$) are used to estimate the outcome from each neuron. The backpropagation learning algorithm is usually used to calculate derivatives of performance of the mean square error with respect to the weight and bias variables.

$$x_1^1 = f\left(\sum_{i=1}^{n} x_i w_{i1}^1 + \theta_1^1\right) \qquad (4)$$

A variety of ANN architectures were implemented, starting with one that contained a number of neurons in the hidden layer equal to the number of attributes used as inputs in each experiment and continuing with topologies resulting from increasing the number of hidden neurons by 1 until it reached the value of sixteen. The initial weights and biases of the network were randomly initialised. Also, the scaled conjugate gradient training function was used which is based on the derivative functions of weights, net inputs and transfer functions. The gradient descent with momentum weight and bias learning function was used for training the ANN.

After experimenting with a variety of architectures and selecting the optimal one, we used Garson's Algorithm [17] to measure the contribution of each independent variable within the best performing ANN. The algorithm partitions the hidden layer weights into components associated with each input node. Next, the percentage of all hidden nodes weights associated with a particular input node was used to measure the relative importance of that attribute.

More specifically, for each input $j$, where $j=1,2,...,i$, the Relative Importance ($RI_j$) was calculated using the following equation:

$$RI_j = \frac{\sum_{m=1}^{Nh}\left[\left[\frac{|w_{jm}^{ih}|}{\sum_{k=1}^{Ni}|w_{km}^{ih}|}\right] \times |w_{jm}^{ho}|\right]}{\sum_{n=1}^{Ni}\left[\sum_{m=1}^{Nh}\left[\left[\frac{|w_{nm}^{ih}|}{\sum_{k=1}^{Ni}|w_{km}^{ih}|}\right] \times |w_{jm}^{ho}|\right]\right]} \quad (5)$$

In equation (5) $N_i$ and $N_h$ are the number of input and hidden neurons respectively, $w$ is the connection weight, the superscripts '$i$','$h$' and '$o$' refer to input, hidden and output layers, respectively and subscripts '$k$', '$m$' and '$n$' refer to input, hidden and input neurons used. In our case there is only one output neuron. According to Garson's algorithm, for each input node $j$ the relative contribution of $j$ to the outgoing signal of each hidden neuron is calculated and converted to a percentage, which serves as a measure of importance for each input node representing each variable. Also, in this algorithm each input that makes the smallest contribution to the final output of the network, as this is reflected through the weight connections, is eliminated. Thus, in each repetition the initial number of variables utilised is lowered gradually by one until half the initial number of variables are left in the dataset.

3.3 Forward Selection and Backward Elimination

The forward selection and backward elimination algorithms were implemented both as Filter (combined with linear Least Squares) and as Wrapper (combined with Ridge Regression) methods. In the case of these two approaches (as well as for the Genetic Algorithm, described in the next subsection) the following evaluation process was followed: As previously mentioned, the data was randomly portioned into percentages of 80% of the samples comprising the training samples and 20% of the samples comprising the testing samples. Then, the training samples were used to select the optimum features with each technique by using a 10-fold cross-validation process. Specifically, we split the training set into 10 parts of almost equal size and we applied the respective FSS technique (with only the selected cost drivers) 10 times, each time evaluating its performance on one part after training on the remaining nine. Performance evaluation included calculating the *Magnitude of Relative Error* (*MRE*) for each fold and the *Mean Magnitude of Relative Error* (*MMRE*) over the whole set (the mean value of the *MRE*s of all projects) at the end of the 10-fold cross-validation process. The full list of performance metrics are explained in detail in section 4.2. The smaller the *MMRE* obtained in the experiments the more optimal the corresponding selected attributes are from the Filter or the Wrapper approach applied in each case.

Both methods (forward selection and backward elimination) select sequentially subsets of features so that the data is best predicted and the *Magnitude of Mean Relative Error* (*MMRE*) is minimised as previously explained. Figure 2 describes these two feature selection approaches. In forward selection we start from an empty feature set and iteratively add a certain feature whose addition to the current subset results in the highest prediction accuracy. This iterative process ends when the addition of any other feature does not result in an improvement. Similarly, in backward elimination, we start from the whole feature set and iteratively remove a certain feature whose removal gives the highest accuracy, again ending the process when any further removals do not result in an improvement.

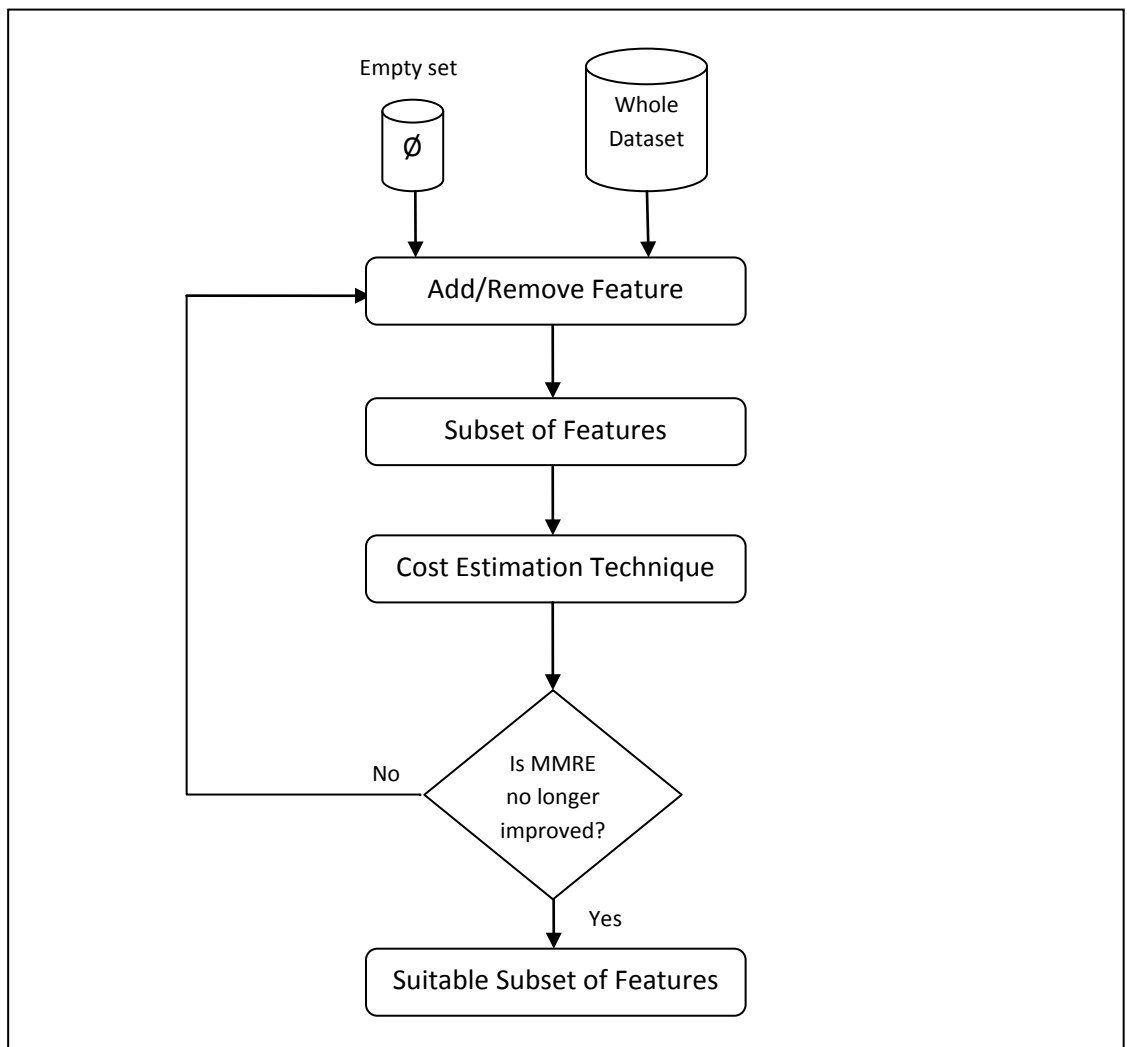

Figure 2: Feature Selection Scheme for Software Cost Estimation

3.4 Genetic Algorithm

The dedicated Genetic Algorithm (GA) for software cost drivers selection was implemented using the MATLAB (R2009a) Genetic Algorithm and Direct Search Toolbox [50] and is an extension of the algorithm proposed in [41]. The algorithm had the following goal: Select the subset of cost drivers that minimises the overall relative error rate obtained with a machine learning technique, in this case Ridge Regression (RR) and Least Squares (LS). The main advantage of using a GA is that it can search the vast space of possible combinations of cost drivers efficiently and reach to a near-to-optimal outcome.

The GA evolves a population of individuals encoded as bit strings of the size of the cost drivers contained in the dataset. The cost drivers represented by the bits set to 1 are taken into account as inputs while all others (set to 0) are not. The individuals at each step, called a generation, are evaluated in the same 10-cross validation way, described in section 3.3. For each individual we assign a score, called its fitness value, indicating how good the solution it represents is. The fitness of each individual defines its likelihood of being selected for the next generation. Until a new generation is completed, individuals from the current generation are selected probabilistically based on their fitness to generate offspring for the new generation. There are also a few individuals, the fittest ones, which are carried to the new generation unchanged, that is, without the application of any genetic operation on them. The same process is repeated until an optimal solution is reached or a stopping criterion is met, which in many cases is a maximum number of generations. The solution represented by the fittest individual in the last population is the one adopted as the resulting solution of the GA.

Firstly, to create each generation the 10% fittest individuals are isolated and placed in the new generation. Then, until the new generation is complete (that is, reaches the maximum size) individuals are selected at a time and recombined and/or mutated to generate new individuals for the new population. The selection of individuals is performed with the stochastic uniform function applied on the rank of the individuals in the current population. In effect this selection function lays out a line in which each individual corresponds to a section of the line of length proportional to its rank. The function then moves along this line in steps of equal size selecting the individual from the section it lands on. The size of the first step is a uniform random number less than the fixed step size.

The crossover operator used the uniform crossover function, with probability of being applied to a selected pair (called crossover rate) being set to 0.8. This function creates a random bit string of the same size as the two parents, called crossover mask, and generates the first child by copying the parts of the first parent at the points where the crossover mask has a 1 and the parts of the second parent at the points where the crossover mask has a 0. For the second child the same process is repeated with the parents reversed. Also, the uniform mutation is used, which flips each bit of an individual with a given probability, which in this case was set to 0.01. The number of generations and individuals in each generation were both set equal to 100.

4. Evaluation Method and Criteria

In this section we briefly describe the dual form RR algorithm, which was the machine learning technique used for obtaining the final predictions, and present the criteria with which these predictions were evaluated.

4.1. Dual form Ridge Regression

Ridge Regression (RR) is an improvement of the classical Least Squares technique, which is one of the dominant methods applied in software cost estimation. RR is a relatively popular regression algorithm that has been applied in a variety of problems, such as ionospheric parameter prediction [38], ecological risk assessment [3] and face recognition [51], and it was also proven successful in producing accurate cost estimations [33, 39]. The main advantages of the RR algorithm include: simplicity in implementation, limited number of free parameters and robustness in the results obtained. In this work we employ the dual variables RR method proposed in [46], which utilises kernel functions to allow the construction of non-linear regressions without having to carry out expensive computations in a high dimensional feature space.

To approximate a set of sample projects $\{(x_1,z_1), ..., (x_n,z_n)\}$, where $x_i \in \Re^n$ is the vector of attributes (cost drivers) for project $i$ and $z_n \in \Re$ is the effort of that project, the RR procedure recommends finding the $w$ which minimises function:

$$a\|w\|^2 + \sum_{i=1}^{n}(z_i - w \cdot x_i)^2, \qquad (6)$$

where $a$ is a positive constant, called the *ridge parameter*. Notice that RR includes Least Squares as a special case (by setting $a = 0$). The RR prediction $\hat{z}_t$ for an input vector $x_t$ is then $\hat{z}_t = w \cdot x_t$.

The dual variables formula, derived in [46], for the prediction of an input vector $x_t$ is:

$$\hat{z}_t = z(K + aI)^{-1}k, \qquad (7)$$

where $z = (z_1, ..., z_n)$ is the vector consisting of the effort outputs of the projects in the training set (the set of known project samples), $K$ is the $n \times n$ matrix of dot products of the input vectors $\{x_1, ..., x_n\}$ of those data samples defined as:

$$K_{i,j} = \mathrm{K}(x_i, x_j), \quad i = 1,...,n, \quad j = 1,....,n \qquad (8)$$

and $k$ is the vector of dot products between $x_t$ and the input vectors of the training samples given as:

$$k_i = \mathrm{K}(x_i, x_t), \quad i = 1,...,n, \qquad (9)$$

and $\mathrm{K}(x, x')$ is the kernel function, which returns the dot product of the vectors $x$ and $x'$ in some feature space.

We used the RBF kernel function, which is the typical choice of kernel in machine learning literature and is defined as:

$$\mathrm{K}(x, x') = \exp(-\frac{\|x - x'\|^2}{2\gamma^2}). \qquad (10)$$

4.2. Evaluation Criteria

The performance accuracy of the approaches investigated was empirically evaluated through a systematic method. The two datasets used were similarly pre-processed, as we will explain later, and the dependent variable in both datasets was the development effort. The performance indicators used in this work were taken by the evaluation metrics commonly used in the software cost estimation literature, that is, the *Magnitude of Mean Relative Error* and the *Prediction Level* at level 0.25. According to [29] these error metrics correspond to the kurtosis of the variable *zeta*, where *zeta=estimate/actual*, or in our case *zeta*= $\hat{z}_i/z_i$.

The *Magnitude of Mean Relative Error* (*MMRE*) is one of the most commonly used criteria for evaluating cost estimation models [14, 47] and is given in equation (11). It aggregates the prediction error for each sample being predicted.

$$MMRE(n) = \frac{1}{n}\sum_{i=1}^{n}\left|\frac{z_i - \hat{z}_i}{z_i}\right| \qquad (11)$$

The *Prediction Level* (*PRED(l)*) defined in equation (12) [11, 48] specifies the fraction of predictions with *Relative Error* (*RE*) (defined in (13)) below a predefined level *l* (*l*=0.25 in this case).

$$PRED(l) = \frac{k}{n} \qquad (12)$$

$$RE(n) = \frac{|z_i - \hat{z}_i|}{z_i} \qquad (13)$$

The FSS approaches that we experimented with were Filters and Wrappers. Specifically, Filter approaches included Backward StepWiseFit (BSWF), Forward StepWiseFit (FSWF), Backward Feature Elimination with Least Squares (LSBFE), Forward Feature Selection with Least Squares (LSFFS), Backward Garson's Algorithm (GARSON) and Genetic Algorithm with Least Squares (LSGA); Wrapper approaches (combined with Ridge Regression) involved Forward Feature Selection (FFS), Backward Feature Elimination (BFE) and finally, Genetic Algorithm (GA). The selected features from each approach are subsequently used as inputs for evaluating the RR algorithm, i.e., to evaluate its cost estimates on both the training and testing sample sets of each of the 10 data partitions.

5. Experiments and Results

This section empirically examines the utilisation of the various FSS techniques on the ISBSG and Desharnais datasets. The *ridge parameter α* and the *kernel gamma γ* were set to 0.1 and 3.5 for the ISBSG dataset, and to 0.05 and 5 for the Desharnais dataset respectively, configurations that were found in previous work to work well when following a fold cross-validation process on the full datasets [39].

5.1 ISBSG Dataset

The ISBSG dataset [23] consists of software project cost records coming from a broad cross section of the industry and range in size, effort, platform, language and development technique. The dataset contains over 1000 samples organised in 92 variables from which only numerical project attributes that are directly related to (and possibly affect) development effort were selected.

5.1.1. Cost Drivers Description and Pre-processing

The ISBSG projects underwent a series of quality checks and pre-processing to create a filtered version of the dataset that would not contain any null values and would conform to the homogeneity and integrity standards set before feeding them as inputs to the models.

More specifically, the following actions of column and row filtering were performed:
a) Excluded attributes that have been measured after the project completion and therefore would not be practical for the cost estimation model we were building.
b) Calculated the number of null records for the remaining attributes and excluded those with more than 40% nulls.
c) Referred to the quality rating provided by reviewers in the ISBSG (the organisation from which the data used in this study is drawn) and excluded project information that were rated higher than grade 'B' (the higher the grade, the lower the quality of the data).
d) Excluded records that contained null values in numerical attributes.
e) Created new binary columns for different values of categorical attributes. In these columns the value of 1 was reported for each project that belonged to the new categories created, and 0 otherwise.
f) All numerical values were normalised to the minimum and maximum range of 0 and 1 respectively so that they would all have the same impact.

The filtered ISBSG dataset produced after pre-processing contained 467 projects and 82 attributes. The attributes selected and used, along with their abbreviations and description, are summarised in Table 1.

**Table 1: Software Cost Attributes (Features) for the ISBSG Dataset**

| ID | CODE | Attribute Name | Short Description |
|---|---|---|---|
| 0 | FCE | Full Cycle Work Effort | Total effort (in hours) recorded against the project. |
| 1-4 | CA | Count Approach | Description of the technique used to size the project. |
| 5 | AFP | Adjusted Function Points | Adjusted function point count number adjusted by the Value Adjustment Factor. |
| 6 | PET | Project Elapsed Time | Total elapsed time for the project (in calendar months). |
| 7 | IY | Implementation Year | Actual year of implementation. |
| 8-11 | DTY | Development Type | Type of development among the categories: new development, enhancement or re-development. |
| 12-23 | OT | Organisation Type | Type of organisation that submitted the project. |
| 24-38 | DT | Development Technique | Techniques used during development. |
| 39-41 | FS | Functional Sizing Technique | Technology to support the functional sizing process. |
| 42-46 | DP | Development Platform | Primary development platform. |
| 47-52 | LT | Language Type | Language type used for the project e.g., 3GL, 4GL. |
| 53-63 | PPL | Primary Programming Language | Primary programming language used for the development e.g., Java, C++. |
| 64-72 | DBS | Database System | Primary technology database used to build or enhance the software. |
| 73-79 | RM | Recording Method | Method used to obtain effort data. |
| 80 | RL | Resource Level | People whose time is included in the work effort data reported. |
| 81 | MTS | Maximum Team Size | Maximum number of people that worked on the project. |
| 82 | ATS | Average Team Size | Average number of people that worked on the project. |

5.1.2. Results

The experimental results on the ISBSG dataset are presented in this section. The minimum, maximum and average *MMRE* and *PRED(0.25)* values obtained are shown in Table 2 for each FSS method. In addition, the last column of the same table (#AT) indicates the number of selected attributes from each method.

Comparing the initial error figures obtained using the full attribute set with the final errors obtained using the optimum feature subset of each method we observe an accuracy decrease in the training phase of the algorithms. As expected, the removal of features in the training phase has a negative impact on performance, whereas on the testing phase this difference is rather insignificant.

**Table 2: Cost Estimation Results Utilising Various FSS on the ISBSG Dataset**

| FSS | RESULTS | TRAINING PHASE | | | | TESTING PHASE | | | | #AT |
| --- | --- | --- | --- | --- | --- | --- | --- | --- | --- | --- |
| | | INITIAL | | FINAL | | INITIAL | | FINAL | | |
| | | MMRE | PRED | MMRE | PRED | MMRE | PRED | MMRE | PRED | |
| BFE | MIN | 0.355 | 0.520 | 0.444 | 0.436 | 0.434 | 0.398 | 0.456 | 0.390 | 34 |
| | MAX | 0.337 | 0.532 | 0.409 | 0.433 | 0.657 | 0.415 | 0.851 | 0.423 | 36 |
| | **AVG** | **0.330** | **0.536** | **0.407** | **0.467** | **0.585** | **0.349** | **0.592** | **0.377** | **32** |
| FFS | MIN | 0.355 | 0.520 | 0.451 | 0.453 | 0.434 | 0.398 | 0.439 | 0.455 | 33 |
| | MAX | 0.337 | 0.532 | 0.418 | 0.456 | 0.657 | 0.415 | 0.771 | 0.463 | 29 |
| | **AVG** | **0.330** | **0.536** | **0.410** | **0.471** | **0.585** | **0.349** | **0.577** | **0.382** | **29** |
| BSWF | MIN | 0.355 | 0.520 | 0.465 | 0.419 | 0.434 | 0.398 | 0.434 | 0.447 | 32 |
| | MAX | 0.313 | 0.523 | 0.414 | 0.436 | 0.763 | 0.309 | 0.880 | 0.407 | 30 |
| | **AVG** | **0.330** | **0.536** | **0.451** | **0.434** | **0.585** | **0.349** | **0.618** | **0.386** | **27** |
| FSWF | MIN | 0.355 | 0.520 | 0.551 | 0.390 | 0.434 | 0.398 | 0.425 | 0.447 | 11 |
| | MAX | 0.313 | 0.523 | 0.482 | 0.401 | 0.763 | 0.309 | 0.817 | 0.382 | 11 |
| | **AVG** | **0.330** | **0.536** | **0.518** | **0.387** | **0.585** | **0.349** | **0.600** | **0.373** | **12** |
| LSBFE | MIN | 0.355 | 0.520 | 0.462 | 0.387 | 0.434 | 0.398 | 0.399 | 0.439 | 39 |
| | MAX | 0.312 | 0.564 | 0.415 | 0.477 | 0.815 | 0.317 | 0.869 | 0.301 | 27 |
| | **AVG** | **0.330** | **0.536** | **0.431** | **0.446** | **0.585** | **0.349** | **0.592** | **0.372** | **32** |
| LSFFS | MIN | 0.355 | 0.520 | 0.461 | 0.448 | 0.434 | 0.398 | 0.447 | 0.407 | 37 |
| | MAX | 0.312 | 0.564 | 0.403 | 0.477 | 0.815 | 0.317 | 0.812 | 0.350 | 31 |
| | **AVG** | **0.330** | **0.536** | **0.442** | **0.435** | **0.585** | **0.349** | **0.596** | **0.363** | **34** |
| GARSON | MIN | 0.355 | 0.520 | 0.566 | 0.384 | 0.434 | 0.398 | 0.478 | 0.325 | 41 |
| | MAX | 0.312 | 0.564 | 0.465 | 0.401 | 0.815 | 0.317 | 0.954 | 0.317 | 41 |
| | **AVG** | **0.330** | **0.536** | **0.489** | **0.402** | **0.585** | **0.349** | **0.686** | **0.320** | **41** |
| LSGA | MIN | 0.355 | 0.520 | 0.457 | 0.410 | 0.434 | 0.398 | 0.432 | 0.407 | 36 |
| | MAX | 0.314 | 0.523 | 0.422 | 0.462 | 0.763 | 0.309 | 0.867 | 0.374 | 29 |
| | **AVG** | **0.330** | **0.536** | **0.434** | **0.436** | **0.585** | **0.349** | **0.607** | **0.355** | **31** |
| GA | MIN | 0.355 | 0.520 | 0.424 | 0.422 | 0.434 | 0.398 | 0.445 | 0.415 | 35 |
| | MAX | 0.312 | 0.564 | 0.374 | 0.491 | 0.815 | 0.317 | 0.748 | 0.382 | 30 |
| | **AVG** | **0.331** | **0.536** | **0.401** | **0.472** | **0.585** | **0.349** | **0.570** | **0.376** | **33** |

The prediction results in the testing phase on average show that the selected features yield comparable results with the full feature sets. This means that a large subset of software cost drivers have a small or insignificant influence in estimating software development effort accurately. Therefore, feature selection for the specific dataset has been found particularly useful for maintaining accuracy and in some cases (i.e., FSWF, LSBFE, GA) even increase performance. The best selection was obtained with the GA feature selection method, among the ones examined, as it yielded the lowest *MMRE* on average in the testing phase. It is also interesting to point out that the FSWF method promoted the smallest in size subset of features among the methods investigated and more importantly, it was able to provide highly accurate development effort estimations, similar to the rest of the FSS methods, which on average selected around less than half the original features. On average from the results obtained from all the techniques 30 optimal features are selected indicating that a quite important attribute reduction has been achieved without any significant compromise on the prediction accuracy of the models.

5.2    Desharnais Dataset

The Desharnais dataset [15] is available from the PROMISE Software Engineering Repository [9] and contains 81 projects developed by a Canadian software house.

5.2.1.    Cost Drivers Description and Pre-processing

The attributes selected and used in this work, along with their abbreviations and description are summarised in Table 3. The number of projects included in the experiments was only 77 as a result of row filtering of null values.

**Table 3: Software Cost Drivers (Features) for the Desharnais**

| ID | CODE | Attribute Name | Short Description |
|---|---|---|---|
| **0** | **EFF** | Effort | Actual development effort (in person-hours). |
| 1 | TE | Team Experience | Team experience (measured in years). |
| 2 | ME | Manager Experience | Manager's experience (measured in years). |
| 3 | LE | Length | Project's duration (measured in months). |
| 4 | TR | Transactions | Count of basic logical transactions in the system. |
| 5 | EN | Entities | Number of entities in the system's data model. |
| 6 | FPA | Function Points Adjusted | Adjusted Function Points count. |
| 7 | ENV | Envergure | Scale of the project. |
| 8 | FPNA | Function Points non Adjusted | Unadjusted Function Points count. |

5.2.2.    Results

Table 4 presents for each FSS approach examined the minimum, maximum and average *MMRE* and *PRED(0.25)* values along with the number of features (#F) selected in the training phase of each algorithm (appearing in the last column).

Commenting briefly on the feature selection techniques applied for the Desharnais dataset the best average prediction rate was obtained with GARSON's filtering. In addition, prediction accuracy was slightly increased using the selected features compared to the full features case in the testing of LSFFS approach, while all other approaches appear to have a decrease in accuracy performance. Overall, examining the error figures using the full attribute set and those using the selected features, we observe that an adequate estimation performance is yielded with only just a few features. This indicates that a small number of features is actually valuable in approximating development effort in the Desharnais dataset, since on average a minimisation of the feature subset size down to more than half the original size performs quite well in both the training and testing phases, even though a minor *MMRE* increase is observed compared to the full (initial) attributes case. Moreover, we observe that on average accuracy performance is at similar levels with the final (selected) attributes and that we obtain more predictions that have *RE* lower than 0.25 which is an indirect indication of performance accuracy increase for the GARSON's FSS (see *PRED* metric). Even though the dimensionality of this dataset is smaller than the ISBSG, the optimisation methods used converge to selecting the same subset of 'appropriate' features and at the same time maintaining prediction accuracy in software cost estimation.

Table 4: Cost Estimation Results Utilising Various FSS on the Desharnais Dataset

| FSS | RESULTS | TRAINING PHASE | | | | TESTING PHASE | | | | #F |
|---|---|---|---|---|---|---|---|---|---|---|
| | | INITIAL | | FINAL | | INITIAL | | FINAL | | |
| | | MMRE | PRED | MMRE | PRED | MMRE | PRED | MMRE | PRED | |
| BFE | MIN | 0.565 | 0.339 | 0.566 | 0.323 | 0.334 | 0.400 | 0.337 | 0.333 | 4 |
| | MAX | 0.482 | 0.419 | 0.489 | 0.435 | 0.815 | 0.267 | 0.916 | 0.267 | 5 |
| | **AVG** | **0.522** | **0.379** | **0.528** | **0.389** | **0.582** | **0.387** | **0.622** | **0.347** | **4** |
| FFS | MIN | 0.565 | 0.339 | 0.566 | 0.323 | 0.334 | 0.400 | 0.337 | 0.333 | 4 |
| | MAX | 0.482 | 0.419 | 0.487 | 0.435 | 0.815 | 0.267 | 0.903 | 0.267 | 6 |
| | **AVG** | **0.522** | **0.379** | **0.528** | **0.395** | **0.582** | **0.387** | **0.622** | **0.340** | **5** |
| BSWF | MIN | 0.565 | 0.339 | 0.584 | 0.435 | 0.334 | 0.400 | 0.345 | 0.400 | 4 |
| | MAX | 0.474 | 0.435 | 0.516 | 0.435 | 0.779 | 0.333 | 0.859 | 0.333 | 2 |
| | **AVG** | **0.522** | **0.379** | **0.552** | **0.427** | **0.582** | **0.387** | **0.623** | **0.413** | **4** |
| FSWF | MIN | 0.556 | 0.323 | 0.623 | 0.387 | 0.387 | 0.467 | 0.416 | 0.400 | 2 |
| | MAX | 0.482 | 0.419 | 0.537 | 0.403 | 0.815 | 0.267 | 0.911 | 0.533 | 3 |
| | **AVG** | **0.522** | **0.379** | **0.577** | **0.427** | **0.582** | **0.387** | **0.653** | **0.413** | **3** |
| LSBFE | MIN | 0.559 | 0.306 | 0.564 | 0.290 | 0.389 | 0.400 | 0.377 | 0.400 | 4 |
| | MAX | 0.448 | 0.435 | 0.462 | 0.403 | 1.264 | 0.067 | 1.170 | 0.200 | 3 |
| | **AVG** | **0.516** | **0.342** | **0.540** | **0.353** | **0.648** | **0.293** | **0.699** | **0.320** | **3** |
| LSFFS | MIN | 0.554 | 0.274 | 0.587 | 0.306 | 0.339 | 0.400 | 0.388 | 0.333 | 2 |
| | MAX | 0.448 | 0.435 | 0.479 | 0.403 | 1.264 | 0.067 | 1.096 | 0.267 | 3 |
| | **AVG** | **0.516** | **0.342** | **0.538** | **0.348** | **0.648** | **0.293** | **0.639** | **0.333** | **3** |
| GARSON | MIN | 0.565 | 0.339 | 0.605 | 0.468 | 0.334 | 0.400 | 0.382 | 0.333 | 4 |
| | MAX | 0.474 | 0.435 | 0.533 | 0.387 | 0.779 | 0.333 | 0.792 | 0.200 | 4 |
| | **AVG** | **0.522** | **0.379** | **0.573** | **0.387** | **0.582** | **0.387** | **0.606** | **0.447** | **4** |
| LSGA | MIN | 0.556 | 0.323 | 0.623 | 0.387 | 0.387 | 0.467 | 0.416 | 0.400 | 2 |
| | MAX | 0.474 | 0.436 | 0.510 | 0.452 | 0.779 | 0.333 | 0.867 | 0.267 | 3 |
| | **AVG** | **0.522** | **0.379** | **0.573** | **0.416** | **0.582** | **0.387** | **0.626** | **0.320** | **3** |
| GA | MIN | 0.566 | 0.339 | 0.566 | 0.323 | 0.334 | 0.400 | 0.337 | 0.333 | 4 |
| | MAX | 0.474 | 0.436 | 0.474 | 0.500 | 0.779 | 0.333 | 0.836 | 0.400 | 5 |
| | **AVG** | **0.522** | **0.379** | **0.530** | **0.377** | **0.582** | **0.387** | **0.614** | **0.327** | **4** |

5.3  Interpretation of the Selected Attributes

A basic objective of this research is to highlight and understand the features that are found more relevant than the rest across the experiments and the various FSS approaches. In this section we present and discuss which attributes were considered more significant, i.e., they globally and consistently appear in the optimal set of features after applying each FSS approach. Table 5 lists in the third and fourth column the features that were selected by all (100%) and by 80% of the runs (see second column) of each method over the 10 different dataset partitions for the ISBSG and Desharnais dataset respectively. In general, we observe that the commonly promoted attributes of all FSS methods present many similarities. This picture, though, is not retained in the 80% case probably due to the fact that more attributes are allowed to enter the pool of significant attributes as the threshold is less strict than the 100% case.

In the ISBSG dataset the attributes promoted in all of the 10 random data splits and all FSS methods, except GARSON's filtering, are the following: AFP, PET, RM and MTS. Additionally, the attributes ATS, LT and DT are also considered important by most of the methods. Even though the features selected in the pre-processing stage comprise the most relevant attributes found in the vast ISBSG dataset, it seems that particularly the size of software and the duration of the project significantly affect effort estimates as they are consistently promoted. These two factors, as expected, drive development effort value and probably if used as inputs in software cost estimation could lead to more accurate results. However, this investigation is something to be addressed in future work. Also, the size of the team contributing to the project is considered quite important. Another element which should be taken into account that relates also with the homogeneity of the dataset itself, is the recording method used to obtain the effort estimates. Finally, we observe that some project-related features, such as language type and development techniques used in the projects, are also attributes with considerable explanatory power over effort since they are among the ones consistently selected.

**Table 5: Selected Features throughout all dataset partitions**

| FSS | Splits % | Ids of Common Features Selected for ISBSG | Ids of Common Features Selected for Desharnais |
|---|---|---|---|
| BFE | 100 % | 5, 6, 23, 41, 76, 81, 82 | none |
| BFE | 80 % | 5, 6, 15, 20, 23, 24, 26, 41, 48, 54, 76, 81, 82 | 3, 5, 7 |
| FFS | 100 % | 5, 6, 24, 49, 76, 81, 82 | 8 |
| FFS | 80 % | 5, 6, 20, 23, 24, 26, 28, 30, 48, 49, 54, 76, 81, 82 | 3, 5, 7, 8 |
| BSWF | 100 % | 1, 2, 5, 6, 76, 81, 82 | 3 |
| BSWF | 80 % | 1, 2, 3, 5, 6, 19, 47, 54, 76, 81, 82 | 3, 4, 8 |
| FSWF | 100 % | 5, 6, 48, 76, 81 | none |
| FSWF | 80 % | 5, 6, 48, 49, 54, 76, 81, 82 | 4, 8 |
| LSBFE | 100 % | 5, 6, 81, 82 | none |
| LSBFE | 80 % | 5, 6, 9, 28, 35, 47, 67, 74, 76, 81, 82 | none |
| LSFFS | 100 % | 5, 6, 49, 76, 81 | none |
| LSFFS | 80 % | 4, 5, 6, 7, 24, 35, 39, 48, 49, 51, 54, 76, 81, 82 | 4, 8 |
| GARSON | 100 % | none | none |
| GARSON | 80 % | 6, 22, 64, 71, 72 | 6 |
| LSGA | 100 % | 5, 6, 48, 76, 81 | none |
| LSGA | 80 % | 4, 5, 6, 28, 35, 39, 48, 51, 54, 65, 76, 81, 82 | 8 |
| GA | 100 % | 5, 6, 24, 26, 54, 76, 81, 82 | none |
| GA | 80 % | 5, 6, 15, 23, 24, 26, 28, 48, 49, 54, 60, 64, 76, 81, 82 | 3, 5, 7 |

In the Desharnais dataset there are no attributes consistently promoted by all random splits (100% case) and the various FSS methods. However, there are a few attributes more consistently promoted than the rest (in 80% of the data splits) and these are FPNA and LE. Also, in some of the random data splits there are a few less frequently promoted attributes, namely TR, EN and ENV. The significant attributes that seem to have direct influence on development effort in the Desharnais case also relate with software size and project length.

As explained below, the results obtained from this work agree with other comparative research studies. Comparing the observations made above with recent related work, in [5] for the ISBSG dataset the commonly selected features by the majority of the algorithms were: DTY, RL and ATS and for the Desharnais all methods consistently selected FPA and development environment. In [32] where the Desharnais dataset was also used, but with different data splits, the most consistently selected features among the majority of the splits and the feature selection methods was the FPNA attribute, which coincides with our findings. Other popular selected features were ME and EN, from which EN is again one of the most commonly promoted features in this work. In [27] the attributes FPA, FPNA, TR and EN were found significantly correlated with the effort values captured in the Desharnais project database and, in addition, case selection could be based on the FPA variable for obtaining even more improved results.

6. Conclusions

Building accurate models for software cost estimation is a tedious process due to the volatile conditions within the developing organisations, such as the complexity of software, change in requirements, personnel turnover, shift of technology, team dynamics etc. as well as the unclear definitions of predictor metrics. Software cost models utilise a wide range of historical project data for approximating development effort, whereas proper thought is rarely given to the selection of suitable project features. The most frequently investigated approaches in literature relate to accurately estimating several fundamental parameters of the development process of a software system, such as the actual size, complexity and duration of the project. However, the majority of the estimation methods proposed neither manage to obtain successful cost forecasts, nor resolve to explicit, measurable and concise set of factors affecting productivity.

Since a plethora of cost drivers affect the evolution of software projects one of the main challenges is to understand and quantify the effect of these project attributes on the effort required to develop software systems. Also, an important attribute of cost drivers is that they relate to project characteristics that can be determined at the initiation phase. In such a case, if the project attributes used in effort estimation were known when the project is at the early stages, then the proposed cost estimation model would be particularly useful and practical. To achieve practicality, several requisites need to co-exist; firstly, the delivery of accurate, transparent and meaningful effort approximations and secondly, the utilisation of measurable and available information at the project initiation stage.

In this work we have tested the accuracy performance of various subsets of cost drivers which were obtained through a range of popular Feature Subset Selection (FSS) algorithms. We have tested a cross-company dataset (ISBSG R9) and a within-company dataset (Desharnais) and identified that FSS approaches are suitable in choosing the more relevant features in both estimation environments and maintaining estimation accuracy. The results of our experiments suggested that the performance of a technique may be maintained at similar levels with only a small subset of the original features. The FSS implementations also examined the appropriateness and efficiency of each subset of selected project attributes for producing accurate effort estimations. Each approach was validated through a number of random validation splits utilised in the experiments. The results showed that the attributes related with the software size, duration, development technique and team size were consistently selected in the optimal subsets of features.

The main contribution of this work is the understanding of the explanatory value of the relationships holding among the various input cost drivers (project features) and the output (development effort), from the empirical software databases used. The comparison of the results obtained showed consistency among the results of the FSS methods, while our findings are in agreement with other related research studies that also identified significant project features on the same datasets. In addition, an important finding is that a rather small number of features is actually required as input for the particular software cost estimation technique investigated to yield successful results. Therefore, the subset of features leads in reducing the model's complexity, training time and without wasting time on needless calculations.

Future plans include performing a thorough sensitivity analysis on the parameters (configurations) of each feature selection approach and investigating how they affect the accuracy of cost estimations, the selected number and type of features. Future work will also investigate the results of combinations of feature selection methods with different evaluation methods, other than Ridge Regression. Another interesting issue is a more thorough examination of the promoted attributes of each dataset and feature selection method. This investigation will attempt to justify why these attributes were selected among the rest, under which assumptions and what are the major differences in the datasets used. Another suggestion for future work is to include statistical analysis of the prediction results obtained in terms of significance and also investigation of the correlations between the features of the subsets selected and their respective effort. Finally, in addition to software cost estimation, the methods examined may be employed in different software engineering models, such as estimating the expected software size and maintenance effort.

References

[1]  Albrecht AJ, Gaffney JR. Software Function Source Lines of Code, and Development Effort Prediction: A Software Science Validation. *IEEE Transactions on Software Engineering* 1983; 9(6): 639-648.

[2]  Albrecht AJ. Measuring Application Development Productivity. *Proceedings of the Joint SHARE, GUIDE, and IBM Application Developments Symposium* 1979; Monterey, California: 83-92.

[3]  Anderson RH, Basta NT. Application of Ridge Regression to Quantify Marginal Effects of Collinear Soil Properties on Phytotoxicity of Arsenic, Cadmium, Lead, and Zinc. *Environmental Toxicology and Chemistry* 2009; 28 (5): USA: 1018–1027.

[4]  Auer M, Trendowicz A, Graser B, Haunschmid E, Biffl S. Optimal Project Feature Weights in Analogy-Based Cost Estimation: Improvement and Limitations. *IEEE Transactions on Software Engineering* 2006; 32 (2): 83-92.

[5]  Azzeh M, Neagu D, Cowling P. Improving Analogy Software Effort Estimation using Fuzzy Feature Subset Selection Algorithm. *Proceedings of 4th International Workshop on Predictor Models in Software Engineering (PROMISE 2008)* 2008; Germany: 71–78.

[6]  Boehm BW. *Software Engineering Economics*. Prentice Hall: 1981.

[7]  Boehm BW, Abts C, Chulani S. Software Development Cost Estimation Approaches – A Survey. *Annals of Software Engineering* 2000; Springer, Netherlands: 10 (1): 177-205.

[8]  Boehm BW, Abts C, Clark B, Devnani-Chulani S. *COCOMO II Model Definition Manual*. The University of Southern California: 1997.

[9]  Boetticher G, Menzies T, Ostrand T. PROMISE Repository of empirical software engineering data http://promisedata.org repository, USA: West Virginia University, Department of Computer Science; 2007.

[10]  Briand LC, Wieczorek I. *Resource Modeling in Software Engineering*. Marciniak J., editor. Encyclopedia of Software Engineering. 2nd edition: Wiley-Interscience: 2002.

[11]  Briand LC, Langley T, Wieczorek I. A Replicated Assessment and Comparison of Common Software Cost Modeling Techniques. *Proceedings of the 22nd international Conference on Software Engineering* 2000; Limerick, Ireland: 377-386.

[12]  Chen Z, Boehm BW, Menzies T, Port D. Finding the Right Data for Software Cost Modeling. *IEEE Software* 2005; 38-46.

[13]  Chen Z, Menzies T, Port D, Boehm BW. Feature Subset Selection Can Improve Software Cost Estimation Accuracy. *Proceedings of the Workshop on Predictor Models in Software Engineering (PROMISE'05)* 2005; St. Louis, Missouri: 1-6.

[14]  Conte SD, Dunsmore HE, Shen YE. *Software Engineering Metrics and Models*. Redwood City, CA: Benjamin-Cummings Publishing Co. Inc.; 1986.

[15]     Desharnais JM. *Analyse Statistique de la Productivite des Projects de Development en Informatique a Partir de la Technique de Points de Fonction*. MSc. Thesis Université du Québec, Montréal: 1989.

[16]     Draper N R, Smith H. *Applied Regression Analysis*. Hoboken, New Jersey: Wiley-Interscience: 1998; 307–312.

[17]     Garson GD. Interpreting Neural-Network Connection Weights. *AI Expert* 1991; 6: 46-51.

[18]     Gray AR, MacDonell SG. Applications of Fuzzy Logic to Software Metric Models for Development Effort Estimation. *Proceedings of the Annual Meeting of the North American Fuzzy Information Processing Society (NAFIPS)* 1997; Syracuse NY, USA: 394-399.

[19]     Gruschke TM, Jørgensen M. The Role of Outcome Feedback in Improving the Uncertainty Assessment of Software Development Effort Estimates. *ACM Transactions on Software Engineering and Methodology* 2008; 17(4), Article 20: 1-35.

[20]     Heiat A. Comparison of Artificial Neural Network and Regression Models for Estimating Software Development Effort. *Information and Software Technology* 2002; 44(15): 911-922.

[21]     Huang SJ, Chiu NH. Optimization of Analogy Weights by Genetic Algorithm for Software Effort Estimation. *Information and Software Technology* 2006; 48: 1034-1045.

[22]     Huang SJ, Chiu NH, Liu YJ. A Comparative Evaluation on the Accuracies of Software Effort Estimates from Clustered Data. *Information and Software Technology* 2008; 50 (9-10): 879-888.

[23]     International Software Benchmarking Standards Group (ISBSG). Victoria: Estimating, Benchmarking & Research Suite Release 9, ISBSG; 2005. Available from: http://www.isbsg.org/

[24]     Jalali O, Menzies T, Baker T, Hihn J. Column Pruning Beats Stratification in Effort Estimation. *3rd International Workshop on Predictor Models in Software Engineering (PROMISE'07: ICSE Workshops)* 2007; Minneapolis, MN, USA: 7-15.

[25]     Jørgensen M, Shepperd MJ. A Systematic Review of Software Development Cost Estimation Studies. *IEEE Transactions on Software Engineering* 2007; 33(1): 33-53. Washington DC: IEEE Computer Press.

[26]     Jun ES, Lee JK. Quasi-optimal Case-selective Neural Network Model for Software Effort Estimation. *Expert Systems with Applications* 2001; 21(1): 1-14. New York: Elsevier.

[27]     Keung JW, Kitchenham BA, Jeffery DR. Analogy-X: Providing Statistical Inference to Analogy-Based Software Cost Estimation. *IEEE Transactions on Software Engineering* 2008; 34(4): 471-484.

[28]     Kirsopp C, Shepperd MJ, Hart J. Search Heuristics, Case-Based Reasoning and Software Project Effort Prediction. *Proceedings of the Genetic and Evolutionary Computation Conference (GECCO'02)* 2002; New York, NY, USA: 1367-1374.

[29]     Kitchenham BA, MacDonell SG, Pickard L, Shepperd MJ. What accuracy statistics really measure. *IEEE Proceedings Software* 2001; 148(3): 81-85.

[30]     Kitchenham BA, Mendes E. Why comparative effort prediction studies may be invalid. *Proceedings of the 5th International Conference on Predictor Models in Software Engineering (PROMISE '09)* 2009; Vancouver, British Columbia, Canada: (4) 1-4.

[31]     Kumar KV, Ravi V, Carr M., Kiran NR. Software Development Cost Estimation Using Wavelet Neural Networks. *Journal of Systems and Software* 2008; 81(11): 1853-1867.

[32]     Li YF, Xie M, Goh TN. A Study of Mutual Information Based Feature Selection for Case Based Reasoning in Software Cost Estimation. *Expert Systems with Applications* 2009; 36 (3): 5921-5931.

[33]     Li YF, Xie M, Goh TN. Adaptive Ridge Regression System for Software Cost Estimating on Multi-Collinear Datasets. *Journal of Systems and Software* 2010 doi:10.1016/j.jss.2010.07.032.

[34]     McCulloch WS, Pitts W. A Logical Calculus of the Ideas Immanent in Nervous Activity. *Bulletin of Mathematical Biology* 1943; 5 (4): 115-133.

[35]     Menzies T, Port D, Chen Z, Hihn J. Specialization and Extrapolation of Software Cost Models. *Proceedings of the 20th IEEE/ACM International Conference on Automated Software Engineering* 2005; Long Beach, California, USA: 384-387.

[36]     Moløkken K, Jørgensen M. A Review of Software Surveys on Software Effort Estimation. *Proceedings of International Symposium on Empirical Software Engineering* 2003; Rome, Italy: 223–230.

[37]     Nguyen V, Steece B, Boehm BW. A Constrained Regression Technique for COCOMO Calibration. *Proceedings of the 2nd ACM-IEEE International Symposium on Empirical Software Engineering and Measurement (ESEM'08)* 2008; Kaiserslautern, Germany: 213-222.

[38]     Papadopoulos H, Haralambous H. Reliable Predictive Intervals for the Critical Frequency of the F2 Ionospheric Layer. *Proceedings of the 19th European Conference on Artificial Intelligence (ECAI'10)* 2010; Frontiers in Artificial Intelligence and Applications 215: 1123-1124. IOS Press.

[39]     Papadopoulos H, Papatheocharous E, Andreou AS. Reliable Confidence Intervals for Software Effort Estimation. *Proceedings of the 2nd Workshop on Artificial Intelligence Techniques in Software Engineering (AISEW'09)* 2009; Thessaloniki, Greece: 211-220.

[40]     Papatheocharous E, Andreou SA. On the Problem of Attribute Selection for Software Cost Estimation: input backward elimination using Artificial Neural Networks. *Artificial Intelligence Applications and Innovations (AIAI'10)* 2010; Larnaca, Cyprus: 287-294.

[41]     Papatheocharous E, Papadopoulos H, Andreou AS. Software Effort Estimation with Ridge Regression and Evolutionary Attribute Selection. *3rd Artificial Intelligence Techniques in Software Engineering Workshop (AISEW'10)* 2010; Larnaca, Cyprus: Available from: http://arxiv.org/abs/1012.5754.

[42]     Port D, Korte M. Comparative studies of the model evaluation criterions MMRE and PRED in software cost estimation research. *Proceedings of the 2nd ACM-IEEE International Symposium on Empirical Software Engineering and Measurement (ESEM'08)* 2008; Kaiserslautern, Germany: 51-60.

[43]     Putnam LH. A General Empirical Solution to the Macro Software Sizing and Estimating Problem. *IEEE Transactions on Software Engineering* 1978; Vol L. SE-4(4): 345-361.

[44]     Putnam LH, Myers W. *Measures for Excellence, Reliable Software on Time, Within Budget*. New Jersey: Yourdan Press: 1992.

[45]     Putnam LH, Myers W. *Five core metrics: the intelligence behind successful software management*. Dorset House Publishing: 2003; ISBN 0-932633-55-2.

[46]     Saunders C, Gammerman A, Vovk V. Transduction with Confidence and Credibility. Proceedings of the 16th International Joint Conference on Artificial Intelligence 1999; Stockholm, Sweden; 2: 722–726. Morgan Kaufmann, Los Altos, CA.

[47]     Shepperd M, Schofield C. Estimating Software Project Effort Using Analogies. *IEEE Transactions on Software Engineering* 1997; 23(12): 733-743.

[48]   Shepperd M, Schofield C, Kitchenham BA. Effort estimation using analogy. *Proceedings of the 18th International Conference on Software Engineering* 1996; Berlin, Germany: 170-178.

[49]   Subramanian GH, Breslawski S. Dimensionality Reduction in Software Development Effort Estimation. *Journal of Systems and Software* 1993; 21(2): 187-196.

[50]   The MathWorks: Genetic Algorithm and Direct Search Toolbox. User's Guide, Version 2.4.2: 2009.

[51]   Xue H, Zhu Y, Chen S. Local Ridge Regression for Face Recognition. *Neurocomputing* 2009; 72(4-6): 1342–1346.